\documentclass[useAMS,usenatbib]{mn2e}
\usepackage{times,graphicx,float}
\usepackage{psfig}

\title[Black-hole fundamental plane from sub-Eddington to quiescent state]
  { Revisit the fundamental plane of black-hole activity from sub-Eddington to quiescent state}
\author[Dong \& Wu]
  {Ai-Jun Dong$^{1,2}$, Qingwen Wu$^1$\thanks{Corresponding author, E-mail: qwwu@hust.edu.cn}  \\
  $^1$School of Physics, Huazhong University of Science and Technology, Wuhan 430074, China\\
  $^2$Department of Electron information Engineering, Chuzhou University, Chuzhou 23900, China\\ }

\pagerange{\pageref{firstpage}--\pageref{lastpage}} \pubyear{0000}

\def\LaTeX{L\kern-.36em\raise.3ex\hbox{a}\kern-.15em
    T\kern-.1667em\lower.7ex\hbox{E}\kern-.125emX}

\begin{document}

\label{firstpage}

\maketitle

\begin{abstract}
  It is very controversial whether radio--X-ray correlation as defined in low-hard state of X-ray binaries (XRBs) can extend to quiescent state (e.g., X-ray luminosity less than a critical value of $L_{\rm X,c} \sim10^{-5.5}L_{\rm Edd}$) or not. In this work, we collect a sample of XRBs and low luminosity active galactic nuclei (LLAGNs) with wide distribution of Eddington ratios ($L_{\rm X}/L_{\rm Edd}\sim10^{-9}-10^{-3}$) to reexplore the fundamental plane between 5 GHz radio luminosity, $L_{\rm R}$, 2-10 keV X-ray luminosity, $L_{\rm X}$, and black hole (BH) mass, $M_{\rm BH}$, namely $\log L_{\rm R}=\xi_{\rm X} \log L_{\rm X}+\xi_{\rm M}\log M_{\rm BH}+\rm constant$. For the whole sample, we confirm the former fundamental plane of Merloni et al. and Falcke et al. that $\xi_{\rm X}\sim 0.6$ and $\xi_{\rm M}\sim 0.8$ even after including more quiescent BHs. The quiescent BHs follow the fundamental plane very well, and, however, FR I radio galaxies follow a steeper track comparing other BH sources. After excluding FR Is, we investigate the fundamental plane for BHs in quiescent state with $L_{\rm X}< L_{\rm X,c}$ and sub-Eddington BHs with $L_{\rm X}> L_{\rm X,c}$ respectively, and both subsamples have a similar slope, $\xi_{\rm X}\sim0.6$, which support that quiescent BHs may behave similar to those in low-hard state. We further select two subsamples of AGNs with BH mass in a narrow range (FR Is with $M_{\rm BH}=10^{8.8\pm0.4}$ and other LLAGNs with $M_{\rm BH}=10^{8.0\pm0.4}$) to simulate the behavior of a single supermassive BH evolving from sub-Eddington to quiescent state. We find that the highly sub-Eddington sources with $L_{\rm X}/L_{\rm Edd}\sim10^{-6}-10^{-9}$ still roughly stay on the extension of radio--X-ray correlation as defined by other sub-Eddington BHs. Our results are consistent with several recent observations in XRBs that the radio--X-ray correlation as defined in low-hard state can extend to highly sub-Eddington quiescent state.

\end{abstract}


\section{Introduction}
   Accreting black holes (BHs) are widely accepted to be the central engines powering most of emission from X-ray binaries (XRBs) and active galactic nuclei (AGNs), where the BH masses are around 3-20 ${M}_{\sun}$ in XRBs and $10^5-10^{10} {M}_{\sun}$ in the center of every large galaxy. The putative intermediate BHs ($10^2-10^{4} {M}_{\sun}$) are still a matter of debate. XRBs are normally transient sources which display complex spectral and timing features during the outbursts, where three main states include high/soft (HS) state, low/hard (LH) state and intermediate state (or steep power-law state). The HS state is characterized by a strong thermal component and a weak power-law component, while the thermal component is weak and power-law component is dominant in LH state. The intermediate state is normally dominated by a steep power-law component (e.g., see \citealt{MR06} and \citealt{zh13} for recent reviews and references therein). It is much complex in AGNs, where statistical investigations suggest that the different types of AGNs can be unified with several parameters (e.g., orientation and radio loudness, \citealt{up95}). Several works have tried to establish connections between the different states of XRBs and different types of AGNs, where low luminosity AGNs (LLAGNs) are analogs of the XRBs in LH state, RQ quasars are analogs of the XRBs in HS state, while AGNs with relativistic jets may correspond to the XRBs in intermediate state (e.g., \citealt{falc04} and \citealt{ko06}).

   The HS state of XRBs and bright AGNs are believed to be powered by a cold, optically thick, geometrically thin standard accretion disc (SSD; \citealt{SS73}) that accompanied with some
   fraction of hot optically thin corona above and below the disc. However, SSD component is normally weak or absent in LH state and most of the radiation comes from the nonthermal power-law component that may come from the hot, optically thin, geometrically thick advection-dominated accretion flows (ADAFs, also called radiatively inefficient accretion flows, RIAFs,
   \citealt{ichi77,nara94,nara95,abra95,wuc06} and see \citealt{yn14} for a recent review and references therein). The anti- and positive correlations between hard X-ray index and Eddington ratio as found
    in both XRBs \citep[e.g.,][]{wugu08} and AGNs \citep[e.g.,][]{wang04,sh08,gc09,con09} may support the transition of accretion modes (e.g., \citealt{cao09,qiao13,cwd14,cw15}).

  The radio and X-ray correlation has long been studied in both XRBs and AGNs, which was used to explore the possible connection between jet and accretion disc \citep[see][for different opinion for the radio emission in quiescent supermassive BHs]{yuan03,lw13}. The quasi-simultaneous radio and X-ray fluxes in LH state of XRBs roughly follow a universal non-linear correlation \citep[$F_{\rm R}\propto F_{\rm X}^{b}$, $b\sim0.5-0.7$,][]{hann98,corb03,gall03,corb13}. Recently, more and more XRBs deviate from the universal correlations \citep[e.g.,][]{xc07,cado07,sole10,jonk10,cori11,ratt12} and form a different `outliers' track with a much steeper radio--X-ray correlation ($b\sim1.4$ as initially found in H1743$-$322, \citealt{cori11}). \citet{cwd14} found that the X-ray spectral evolutions are different for the data points in the universal and `outliers' tracks, which support that these two tracks may be regulated by radiatively inefficient and radiatively efficient accretion discs respectively \citep[see also][]{cori11,hw14,qiao15}. It is the similar case in AGNs, where LLAGNs follow a shallower radio-X-ray correlation \citep[e.g., the index $b\sim0.6$,][]{wu13} while bright AGNs normally follow a much steeper correlation \citep[e.g., $b\sim1.6,$][]{dw14,pa15}.

  By taking into account the BH mass, the universal radio--X-ray correlation of $F_{\rm R}\propto F_{\rm X}^{\sim 0.6}$ was extended to AGNs, which is called ``fundamental plane'' of BH activity \citep[e.g.,][]{merl03},
    \begin{equation}
   \log L_{\rm R}=0.60^{+0.11}_{-0.11}\log L_{\rm X}+0.78^{+0.11}_{-0.09}\log M_{\rm BH}+7.33^{+4.05}_{-4.07},
   \end{equation}
   where $L_{\rm R}$ is 5 GHz nuclear radio luminosity in unit of $\rm erg\ s^{-1}$, $L_{\rm X}$ is 2-10 keV nuclear X-ray luminosity in unit of $\rm erg\ s^{-1}$, and $M_{\rm BH}$ is the BH mass in unit of ${M}_{\sun}$ \citep[see also][]{falc04,wangy06,kord06,li08,yuan09,gult09a,plot12}. This fundamental plane is tightest for LH state of XRBs and sub-Eddington AGNs \citep[][]{kord06}. Both `outliers' of XRBs and bright AGNs follow a steeper radio--X-ray correlation and a positive hard X-ray photon index--Eddington ratio correlation ($\Gamma-L/L_{\rm Edd}$), which is most possibly regulated by disk-corona model \citep[][]{dw14}. Based on these similarities, \citet{dw14} proposed a new fundamental plane for radiatively efficient BHs,
  \begin{equation}
  \log L_{\rm R}=1.59^{+0.28}_{-0.22} \log L_{\rm X}- 0.22^{+0.19}_{-0.20}\log M_{\rm BH}-28.97^{+0.45}_{-0.45}.
  \end{equation}
  These two universal correlations of \citet{merl03} and \citet{dw14} with much different slopes of $\xi_{X}$ are most possibly regulated by radiatively inefficient and radiatively efficient BH sources respectively.

 The nature of BHs in quiescent state remains an open issue. The anti-correlation between hard X-ray photon index and Eddington ratio ($\Gamma-L_{\rm X}/L_{\rm Edd}$) are found for LH-state of BHs with $L_{\rm X}/L_{\rm Edd}\la0.1\%$ \citep[e.g.,][]{wugu08}. However, this anti-correlation as found in LH state does not continue once the XRB enters quiescence, where $\Gamma$ keeps roughly a constant when $L_{\rm X}/L_{\rm Edd}\la10^{-5}$ \citep[e.g.,][see also Yang et al. 2015 for possible evidence in AGNs]{plot13}. The physical reason is still unclear, where \citet{yang15} proposed that the X-ray emission in quiescent state may be dominated by the jet and the value of $\Gamma$ should keep as a constant while the X-ray emission dominantly come from ADAF in LH state.  The spectral energy distribution (SED) modeling also preferred the pure jet model for these quiescent BHs \citep[e.g.,][]{xie14,plot15}. \citet{yc05} explored the universal correlation of $F_{\rm R}\propto F_{\rm X}^{0.7}$ as found in LH state of XRBs based on ADAF-jet model and predicted that the radio--X-ray correlation will also deviate from that of LH state and will become steeper as $F_{\rm R}\propto F_{\rm X}^{1.23}$ when the X-ray luminosity is lower than a critical luminosity ($L_{\rm X,c}\sim 10^{-6}-10^{-5} L_{\rm Edd}$), where both the radio and X-ray emission should dominantly come from the jet. \citet{yuan09}'s results seemed to support this prediction based on a small sample of LLAGNs with X-ray luminosity roughly below the critical value. However, several quiescent BH XRBs seem to challenge this prediction that they still follow the radio--X-ray correlation as defined in LH state very well and do not evidently deviate even for $L_{\rm X}< 10^{-8}L_{\rm Edd}$ (\citealt[see][for A0620-00 and XTE J1118+480]{gall06,gall14} and see also \citealt{car10}). Therefore, it is still unclear whether quiescent XRBs follow the radio--X-ray correlation as found in LH state of XRBs or not.


 In recent years, more and more radio and X-ray observations were available for quiescent supermassive BHs in LLAGNs and the data also increased for XRBs in quiescent state. In this work, we aim to reexplore the radio-X-ray correlation and the fundamental plane for BH sources from sub-Eddington down to quiescent state by collecting more quiescent BHs. Throughout this work, we assume the following cosmology for AGNs: $H_{0}=70\ \rm km\ s^{-1} Mpc^{-1}$, $\Omega_{0}=0.27$ and $\Omega_{\Lambda}=0.73$.

\begin{table*}
\begin{minipage}{100mm}
\footnotesize
\centerline{\bf Table 1. The data of XRBs in quiescent state. }
\tabcolsep 1.0mm
\begin{tabular}{lccccccl}\hline\hline

Name  & $d_{L}$ & $L_{\rm{X}}^{\rm 2-10 keV}$        &$L_{\rm{R}}^{\rm 5GHz}$         &$M_{\rm{BH}}$     & $\frac{L_{\rm X}}{L_{\rm{Edd}}}$ & $\frac{L_{\rm X,c}}{L_{\rm{Edd}}}$  & Ref.$^{a}$\\
      &kpc      &$\log(\rm ergs/s)$   &$\log(\rm ergs/s)$   &$\log(M_{\sun})$  \\
\hline
XTE J1752-223 &3.5  &32.80 &27.71&0.99 &-6.30 &-5.52& 1, 2, 2, 1\\
H1743-322     &7.5  &33.00 &28.35&1.12 &-6.23 &-5.55& 3, 4, 4, 5\\
XTE J1118+480 &1.7  &30.52 &25.92&0.88 &-8.47 &-5.51& 6, 7, 7, 7\\
A0620-00      &1.2  &30.30 &26.85&0.82 &-8.63 &-5.50& 6, 8, 8, 9\\
V404 Cyg      &7.5  &31.98 &27.97&1.08 &-7.21 &-5.54& 10, 8, 8, 9\\
V404 Cyg      &7.5  &33.69 &28.75&1.08 &-5.50 &-5.54& 10, 8, 8, 9\\
\hline
\end{tabular}
\end{minipage}

\begin{minipage}{170mm}
$^a$ The reference for distance, X-ray luminosity, radio luminosity and BH mass respectively, which are shown as follows: 1)\citet{shap10}; 2)\citet{ratt12}; 3)\citet{jonk10}; 4)\citet{cori11}; 5)\citet{ru13}; 6)\citet{ru06}; 7)\cite{gall14}; 8)\citet{fen10}; 9)\citet{zh13}; 10)\citet{mj09}.
\end{minipage}
\
\end{table*}

\begin{table*}
\centering
\begin{minipage}{180mm}
\footnotesize
  \centerline{\bf Table 2. The data of LLAGNs. }
\tabcolsep 1.20mm
\begin{tabular}{lcccccclcccccccl}\hline\hline
Name  & $L_{\rm{X}}^{\rm 2-10 keV}$ &$L_{\rm{R}}^{\rm 5GHz}$       & $M_{\rm{BH}}$& Refs.& $\frac{L_{\rm X}}{L_{\rm{Edd}}}$& $\frac{L_{\rm X, c}}{L_{\rm{Edd}}}$&
Name  & $L_{\rm{X}}^{\rm 2-10 keV}$ &$L_{\rm{R}}^{\rm 5GHz}$       & $M_{\rm{BH}}$&Refs.& $\frac{L_{\rm X}}{L_{\rm{Edd}}}$ & $\frac{L_{\rm X, c}}{L_{\rm{Edd}}}$ \\
&$\log (\rm ergs/s)$ &$\log (\rm ergs/s)$&$\log M_{\sun}$&& & & &$\log (\rm ergs/s)$&$\log (\rm ergs/s)$&$\log (M_{\sun})$&  \\
\hline
         &       &       &      &       &       &$L_{\rm X} \ga L_{\rm X,c}$\\
\hline

NGC 266	    &40.88	&37.95 	&8.37 	    &1,2,3      &-5.60 	&-6.78              &NGC 4203	&39.69	&36.70 	&7.79 	&1,2,3      &-6.21 	&-6.68\\
NGC 2768	&39.46$^{a}$&37.39 	&7.94   &4,2,3      &-6.12 	&-6.71              &NGC 4235	&42.25	&37.62 	&7.60 	&1,5,3      &-3.46 	&-6.65\\
NGC 3031    &39.38	&36.03 	&7.73 	    &1,5,3      &-6.46 	&-6.67              &NGC 4258	&40.89	&35.78 	&7.57 	&1,5,3      &-4.79 	&-6.64\\
NGC 3147	&41.87	&37.91 	&8.29 	    &1,2,3      &-4.53 	&-6.77              &NGC 4395	&39.58	&34.59 	&4.63 	&1,2,3      &-3.16 	&-6.14\\
NGC 3169	&41.05	&37.19 	&8.01 	    &1,2,3      &-5.07 	&-6.72              &NGC 4450	&40.02	&36.78	&7.40 	&1,2,3      &-5.49 	&-6.61\\
NGC 3226	&39.99	&37.20 	&8.06   	&1,2,3      &-6.18 	&-6.73              &NGC 4477	&39.60	&35.64 	&7.89 	&1,5,3      &-6.40 	&-6.70\\
NGC 3227	&41.70	&36.31 	&7.41    	&1,2,3      &-3.82 	&-6.62              &NGC 4548	&39.74	&36.55$^{c}$ 	&7.08 	&1,10,3      &-5.45 	&-6.56\\
NGC 3516	&42.39	&37.28 	&7.94   	&1,2,3      &-3.66 	&-6.71              &NGC 4565	&39.56	&36.26 	&7.41 	&1,2,3      &-5.96 	&-6.62\\
NGC 3718	&40.44	&36.96	&7.69   	&1,6,3      &-5.36 	&-6.66              &NGC 4579	&41.15	&37.55 	&7.77 	&1,2,3      &-4.73 	&-6.68\\
NGC 3884	&41.89	&37.94$^{b}$ &8.19 	&1,7,3      &-4.41 	&-6.75              &NGC 4594	&40.69	&37.85 	&8.46 	&1,11,3      &-5.88 	&-6.79\\
NGC 3941	&39.27	&35.61 	&7.37   	&1,5,3      &-6.21 	&-6.61              &NGC 4639	&40.18	&35.40 	&6.77 	&1,5,3      &-4.70 	&-6.51\\
NGC 3998	&41.57	&38.36 	&8.89   	&8,2,3      &-5.43 	&-6.87              &NGC 4772	&39.30	&36.48 	&7.57 	&1,2,3      &-6.38 	&-6.64\\
NGC 4138	&40.11	&36.13 	&7.19   	&1,5,3      &-5.19 	&-6.58              &NGC 5033	&40.70	&36.94 	&7.60 	&1,5,3      &-5.01 	&-6.65\\
NGC 4143	&40.03	&37.18 	&8.16   	&1,2,3      &-6.24 	&-6.74              &NGC 5548	&43.23	&37.89 	&8.81 	&1,5,3      &-3.69 	&-6.85\\
NGC 4168	&39.87	&36.63	&7.97	    &9,2,3     &-6.21 	&-6.71              &NGC 7626	&40.97	&38.48 	&8.71 	&1,2,3      &-5.85 	&-6.84\\

\hline
         &       &       &      &       &       &$L_{\rm X} \la L_{\rm X,c}$\\
\hline
NGC 404	    &37.02	&33.50 	&5.16 	&1,12,5     &-6.25 	&-6.23  &NGC 4459   	&38.87	&36.09 	&7.82 	&1,11,3  &-7.06 	&-6.69\\
NGC 821	    &38.30	&35.40 	&8.21 	&13,13,3   &-6.90 	&-6.75  &NGC 4501   	&38.89	&36.28 	&7.79 	&1,5,3  &-7.01 	&-6.68\\
NGC 2787	&38.79	&37.01 	&8.14 	&1,2,3     &-7.46 	&-6.74  &NGC 4552   	&39.49	&38.35 	&8.55 	&1,6,3  &-7.17 	&-6.81\\
NGC 2841	&38.26	&36.00$^{c}$ 	&8.31 	&1,10,3     &-8.16 	&-6.77  &NGC 4636   	&39.38	&36.76$^{c}$ 	&8.14 	&1,10,3  &-6.87 	&-6.74\\
NGC 3245	&39.29	&36.98 	&8.21 	&1,11,3     &-7.03 	&-6.75  &NGC 4649   	&38.10	&37.48$^{d}$ 	&9.07 	&1,11,3  &-9.08 	&-6.90\\
NGC 3379	&37.53	&35.73 	&8.18 	&1,11,3     &-8.76 	&-6.75  &NGC 4698   	&38.69	&35.59 	&7.57 	&1,5,3  &-6.99 	&-6.64\\
NGC 3607	&38.63	&37.01 	&8.40 	&1,14,3    &-7.88 	&-6.78  &NGC 4736   	&38.48	&35.51$^{c}$ 	&7.05 	&1,10,3  &-6.68 	&-6.55\\
NGC 3627	&37.60	&36.11	&7.24	&12,15,3    &-7.81 	&-6.60  &NGC 4762       &38.26  &36.58$^{c}$  &7.63   &1,10,3  &-7.48  &-6.65\\
NGC 3628    &38.24  &36.13$^{c}$  &7.24   &1,10,3     &-7.11   &-6.59  &NGC 5846   	&39.65	&36.73 	&8.43 	&1,2,3  &-6.89 	&-6.79\\
NGC 4216	&38.91	&36.58$^{c}$ 	&8.09 	&1,10,3     &-7.29 	&-6.73  &NGC 5866   	&38.60	&37.07	&7.81 	&1,2,3  &-7.32 	&-6.68\\
NGC 4278	&39.64	&37.95 	&8.61 	&1,6,3     &-7.08 	&-6.82  &Sgr A$^{*}$    &33.34	&32.50	&6.41$^e$&16,16,16 &-11.18 &-6.45\\

\hline\hline
\end{tabular}
\end{minipage}

\begin{minipage}{180mm}
Note:
 $a$) The 0.3-7keV X-ray flux is converted to 2-10keV flux by assuming a power law spectrum with $\Gamma=2$ (see also, \citet{mill12}) \\
 $b$) The 5 GHz radio core emission of NGC 3884 is observed by MERLIN, operated by Jodrell Bank Observatory with resolution of $\sim 0.5^{''}$;\\
 $c$) The 5 GHz radio luminosity is extrapolated from 15 GHz by assuming $f_{\nu} \propto \nu^{-0.5}$ (e.g., \citealt{hu01} and \citealt{ho02});\\
 $d$) The 5 GHz radio core emission of NGC 4649 is extrapolated from 1.5 GHz;\\
 $e$) The BH mass of Sgr A$^{*}$ is estimated from stellar kinematics;\\
 References for radio luminosity, X-ray luminosity and BH mass: 1) \citet{ho09}; 2) \citet{nagar05}; 3) \citet{hg09}; 4) \citet{bor11};
 5) \citet{hu01}; 6) \citet{nagar01}; 7) \citet{filho06}; 8) \citet{youn11}; 9) \citet{pa06}; 10) \citet{nagar02}; 11) \citet{ho02}; 12) \citet{yuan09}; 13)\citet{pe07}; 14)\citet{fab89}; 15) \citet{lm97};  16) \citet{merl03};  \\

\end{minipage}

\end{table*}

   \section{Sample}
   For purpose of our work, we select the XRBs and LLAGNs from sub-Eddington to quiescent state, where we particularly include much more quiescent BHs compared former works . \citet{yc05} predicted that the X-ray emission should be dominated by jet and the radio--X-ray correlation will become steeper if the X-ray luminosity of BH systems is lower than a critical value through modeling the radio--X-ray correlation of XRBs in LH state with the ADAF-jet model. This critical X-ray luminosity is
    \begin{equation}
  \log \frac{L_{\rm X,c}}{L_{\rm Edd}}=-5.356-0.17\log\frac{M_{\rm BH}}{M_{\sun}},
   \end{equation}
    where the critical Eddington ratio is also roughly consistent with the change of hard X-ray spectral evolution from LH to quiescent state in XRBs \citep[e.g.,$L_{\rm X} \la 10^{-5.5}L_{\rm Edd}$,][]{plot13}. To separate the quiescent BHs from our samples, we simply use the criteria of equation (3).

   For XRBs, we select three sources with fruitful simultaneous or quasi-simultaneous radio and X-ray observations with $L_{\rm X}\la10^{-3}L_{\rm Edd}$ in LH state (GX 339-4, \citealt{cwd14}; XTE J1118+480 and V404 Cyg, \citealt{fen10} and references therein). Some LH-state XRBs that stay in `outliers' track or have only few simultaneous observations are neglected. Five quiescent XRBs with simultaneous or quasi-simultaneous radio and X-ray observations were selected from literatures, which are XTE J1752-223\citep[][]{ratt12}, H1743-322 \citep[][]{cori11}, XTE J1118+480 \citep[][]{gall14}, A06200-00 and V404 Cyg \citep[][]{fen10}. The radio luminosity at 5 GHz and X-ray luminosity in 2-10 keV band were adopted in our work, where the radio emission observed in different waveband is extrapolated to 5 GHz assuming a typical radio spectral index of $\alpha=-0.12$ \citep[$F_{\nu}\propto\nu^{-\alpha}$, e.g.,][]{corb13} and it is the same case for X-ray luminosity by assuming a typical photon index of $\Gamma=1.6$ in LH state. The distance, BH mass, X-ray luminosity, radio luminosity and critical Eddington ratio of quiescent XRB are reported in Table 1, where radio and X-ray luminosities for LH state of XRBs can be found in above references.

\begin{table*}
\centering
\begin{minipage}{150mm}
\footnotesize
\centerline{\bf Table 3. The data of FR Is.}
\tabcolsep 1.0mm
\begin{tabular}{lcccccclclcccc}\hline\hline

Name  & $L_{\rm{X}}^{\rm 2-10 keV}$ &$L_{\rm{R}}^{\rm 5GHz}$ &$M_{\rm{BH}}$ &Refs.  & $\frac{L_{\rm X}}{L_{\rm{Edd}}}$&$\frac{L_{\rm X,c}}{L_{\rm{Edd}}}$&
Name  & $L_{\rm{X}}^{\rm 2-10 keV}$ &$L_{\rm{R}}^{\rm 5GHz}$ &$M_{\rm{BH}}$ &Refs.  & $\frac{L_{\rm X}}{L_{\rm{Edd}}}$&$\frac{L_{\rm X,c}}{L_{\rm{Edd}}}$\\
      &$\log\rm (ergs/s)$        &$\log\rm (ergs/s)$             &$\log(M_{\rm \odot})$   &       &                 &                 &
      &$\log\rm (ergs/s)$        &$\log\rm (ergs/s)$             &$\log(M_{\rm \odot})$   &      \\
\hline
      &              &                   &              &       &                &                  &$L_{\rm X} \ga L_{\rm X,c}$\\
\hline
3C 31      &40.67 &39.45  &8.70 &1,1,2 &-6.14   &-6.84  &3C 442A    &41.10 &38.21  &8.40 &1,1,6 &-5.41    &-6.78\\
3C 66B     &41.10 &39.97  &8.84 &1,1,2 &-5.85   &-6.86  &3C 449     &40.35 &39.08  &8.54 &1,1,2 &-6.30    &-6.81\\
3C 76.1    &41.28 &39.07  &8.08 &1,1,3 &-4.95   &-6.74  &3C 465     &41.04 &40.41  &9.13 &1,1,2 &-6.20    &-6.91\\
3C 83.1B   &41.13 &39.46  &9.01 &1,1,2 &-5.99   &-6.89  &NGC 315    &41.63 &40.41  &8.89 &7,8,9 &-5.37    &-6.87\\
3C 84      &42.91 &42.32  &8.64 &1,4,2 &-3.84   &-6.82  &NGC 507    &40.66 &37.67  &8.91 &7,10,9 &-7.12   &-6.87\\
3C 264     &41.87 &39.98  &8.61 &1,5,2 &-5.04   &-6.85  &NGC 1052   &41.53 &39.85  &8.25 &7,11,9 &-4.83   &-6.76\\
3C 296     &41.49 &39.68  &8.80 &1,1,2 &-5.42   &-6.85  &NGC 4261   &40.59 &39.21  &8.92 &7,8,9 &-6.44    &-6.87\\
3C 305     &41.42 &39.75  &8.10 &1,1,2 &-4.79   &-6.73  &NGC 6109   &40.35 &39.44  &8.56 &1,1,2 &-6.31    &-6.81\\
3C 338     &40.31 &40.03  &8.92 &1,1,2 &-6.73   &-6.87  &NGC 6251   &41.60 &40.35$^{a}$  &8.97 &6,1,2 &-5.48    &-6.88\\
3C 346     &43.40 &41.83  &8.89 &1,1,2 &-3.60   &-6.87\\
\hline
      &              &                   &              &       &                &                  &$L_{\rm X} \la L_{\rm X,c}$\\
\hline
3C272.1   &39.35 &38.22  &8.80 &1,1,2 &-7.56   &-6.85  &3C274     &40.59 &39.87  &9.48 &1,1,2 &-7.00   &-6.97\\
\hline\hline
\end{tabular}
\end{minipage}

\begin{minipage}{150mm}
 Note:$a$: the radio luminosity is derived from the observation of VLBI. \\
 References: 1) \citet{hard09}; 2) \citet{wu11}; 3) \citet{woo02}; 4) \citet{lm97}; 5) \citet{lar04}; 6) \citet{wu13}; 7) \citet{ho09}; 8)\citet{merl03};
 9) \citet{hg09}; 10) \citet{mu11}; 11) \citet{fab89}

\end{minipage}
\end{table*}

  For supermassive BH sources, we select a sample from a Palomar Survey of nearby galaxies, which is a magnitude-limited spectroscopic study of a nearly complete sample of 486 bright ($B_{T}\leq$12.5 mag) northern ($\delta>0^{\rm o}$) galaxies (see Ho et al. 1997a,b for more details, and references therein). The nuclear X-ray luminosities and central stellar velocity dispersions for the sources in this survey are further given in \citet{ho09} and \citet{hg09} respectively, where the sources observed by $Chandra$ and/or $XMM-Newton$ are selected in this work. For our purpose, we exclude the sources with only upper limit of X-ray luminosity and the sources with $L_{\rm X}>10^{-3}L_{\rm Edd}$ which may stay in radiatively efficient phase \citep[e.g.,][]{ho08}. The radio core emission of these sources is selected \citep[e.g.,][]{hu01,nagar01,nagar05,filho06}, where most of sources are observed by Very Large Array (VLA, at resolution of $\sim 1^{"}$) or even Very Long Baseline Array (VLBA) at higher resolution and NGC 3884 observed by MERLIN is also selected. The radio fluxes of 7 sources observed by VLA at 15 GHz are converted to 5 GHz by assuming $F_{\nu}\propto\nu^{-\alpha}$ \citep[$\alpha=0.5$ is adopted, e.g.,][]{hp01,ho02}. The fifteen putative Compton-thick sources are also neglected (NGC 1068, NGC 676, NGC 1167, NGC 1667, NGC 2273, NGC 3185, NGC 3489, NGC 3982, NGC 5194, NGC 7743 (\citealt{pa06}) Mrk 3, NGC 4945, NGC 7479 \citep{geo11} and  NGC 2655, NGC 2639 (\citealt{ter05}), since that the X-ray and other band emission may be seriously obscured. The BH mass of selected sources is calculated from the $M_{\rm BH}-\sigma_{*}$ relation of \citet{gult09b},
   \begin{equation}
  \log \frac{M_{\rm BH}}{M_{\sun}}=(8.12\pm0.08)+(4.24\pm0.41)\log \frac{\sigma_{*}}{200kms^{-1}}
  \end{equation}
  where $\sigma_{*}$ is selected from \citet{hg09}. The famous supermassive BH in our Galactic center (Sgr A*, the quiescent state) is also included even it was not included in Palomar sample. We note that Palomar sample include several traditionally radio-loud sources with relativistic large-scale jets (e.g., NGC 315, NGC 507, NGC 1275, NGC 4261, NGC 4374 and NGC 4486), where the central engine may be different from those without relativistic jets. To explore the radio--X-ray correlation in these strong radio sources and compare it with that of \citet{yuan09}, we select 17 more FR Is with relativistic jets from 3CRR sample. Similar to LLAGN sample, the selected FR Is should be observed by $Chandra/XMM-Newton$ in X-ray band and VLA/VLBA at radio waveband. The radio core emission of these FR Is is adopted in this work. The BH mass are also calculated from their central stellar velocity dispersions \citep[][]{ho09,wu11}. In total, we select 52 LLAGNs without evident relativistic large-scale jets (22 sources have $L_{\rm X}<L_{\rm X,c}$) and 21 FR Is with strong jets (2 sources have $L_{\rm X}<L_{\rm X,c}$), which are listed in Tables (2) and (3) respectively, where the X-ray luminosity, radio luminosity and BH mass for each selected source are reported.

   \begin{figure*}
   \includegraphics[width=13cm]{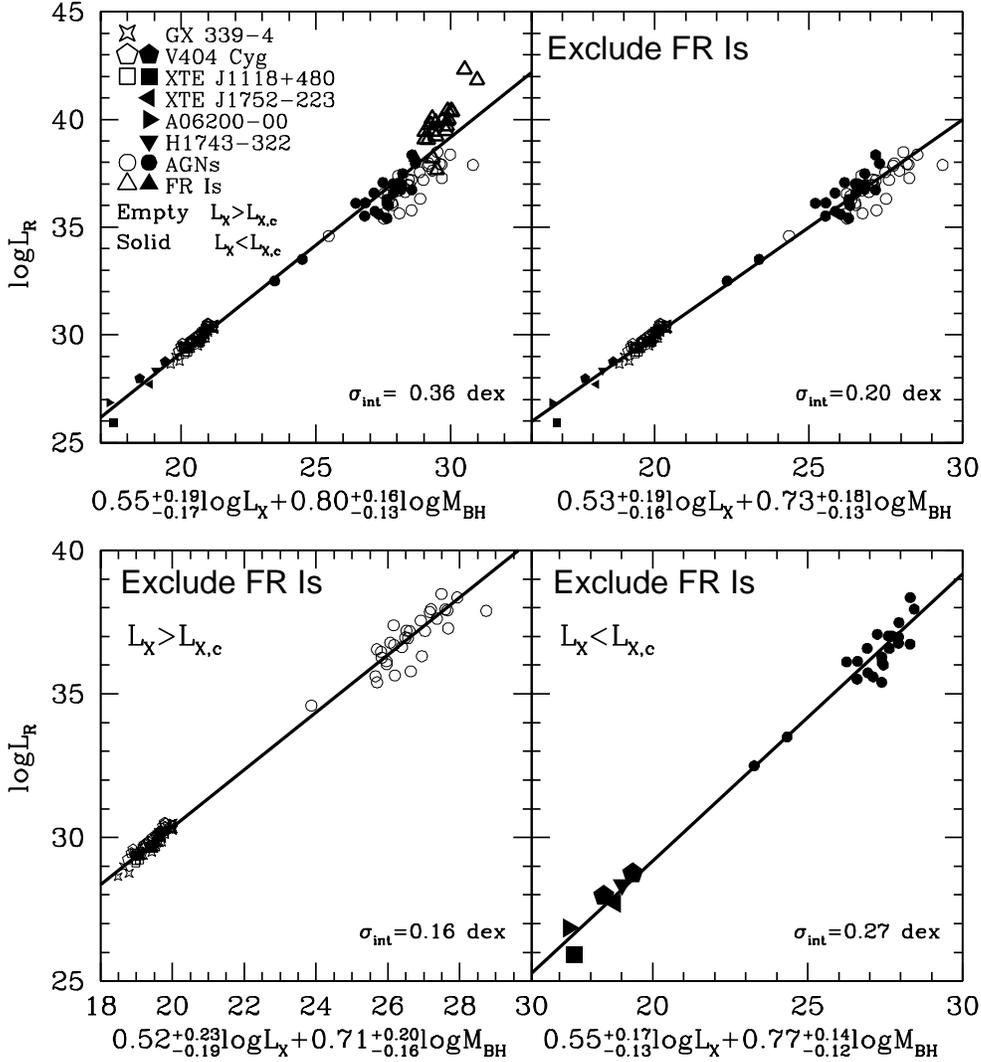}
   \caption{The fundamental plane for the BH activities, where the empty and solid points represent the sources with $L_{\rm X}< L_{\rm X,c}$ and $L_{\rm X}>L_{\rm X,c}$ respectively. Top-left panel represent the plane for all BH sources as selected in our sample; Top-right panel represent the plane for BH sources after excluding FR Is; Bottom panels represent the plane for BHs (exclude FR Is) with $L_{\rm X}>L_{\rm X,c}$ (left) and $L_{\rm X}< L_{\rm X,c}$ (right) respectively. The solid lines are the best fittings. }
  \end{figure*}

\section{Method and Result}

  To explore the fundamental plane for the sub-Eddington and quiescent BHs, we take the form of $\log L_{\rm R}=\xi_{\rm X} \log L_{\rm X}+\xi_{\rm M}\log M_{\rm BH}+c_{0}$ as in \citet{merl03}, where $L_{\rm R}$ is 5 GHz radio luminosity, $L_{\rm X}$ is 2-10 keV X-ray luminosity.  To find the multi-parameter relation, we adopt a similar approach as that of \citet{merl03} and minimize the following statistic,
 \begin{equation}
  \chi^{2}=\sum\limits_{i} \frac{(y_i-c_0-\xi_{\rm X}X_i-\xi_{\rm M}M_i)^2}{\sigma^2_{\rm R}+\xi_{\rm X}^2\sigma^2_{\rm X}+\xi_{\rm M}^2\sigma^2_{\rm M}},
   \end{equation}
  where $y_i=\log L_{\rm R}$, $X_i=\log L_{\rm X}$, $M_i=\log M_{\rm BH}$ and $c_0$ is a constant. Instead of assuming the isotropic uncertainties with $\sigma_{L_{\rm R}}=\sigma_{L_{\rm X}}=\sigma_{\rm M}$ as in \citet{merl03}, we adopt the typical observational uncertainties $\sigma_{L_{\rm R}}=0.2$ dex \citep[e.g.,][]{hp01}, $\sigma_{L_{\rm X}}=0.3$ dex \citep[e.g.,][]{st05}, and $\sigma_{\rm M}=0.4$ dex \citep[e.g.,][]{vp06} for AGNs and the typical variations (within one day) $\sigma_{L_{\rm R}}=0.1$ dex, $\sigma_{L_{\rm X}}=0.15$ dex\citep[e.g.,][]{cori11,corb13}, and typical uncertainty of BH mass $\sigma_{\rm M}=0.15$ dex \citep[e.g.,][]{zh13} for XRBs.

  In top-left panel of Figure 1, we present the fundamental plane for all selected BH sources from sub-Eddington to quiescent state. The best fit for the whole sample is
  \begin{equation}
  \log L_{\rm R}=0.55^{+0.19}_{-0.17} \log L_{\rm X}+ 0.80^{+0.16}_{-0.13}\log M_{\rm BH}+9.17^{+0.34}_{-0.34},
   \end{equation}
  with an intrinsic scatter of $\sigma_{\rm int}=0.36$ dex. From this panel, we find (1) the quiescent BHs still roughly follow the correlation as defined by the whole sample and do not show evident deviation; (2) the FR Is seem to follow a steeper track comparing with other sources. After excluding the FR Is from the whole sample, we present the fundamental plane for other sources in top-right panel of Figure 1 and the best fit is

  \begin{equation}
  \log L_{\rm R}=0.52^{+0.23}_{-0.19} \log L_{\rm X}+ 0.73^{+0.18}_{-0.13}\log M_{\rm BH}+9.97^{+0.31}_{-0.30},
   \end{equation}
  with an intrinsic $\sigma_{\rm int}=0.20$ dex. It can be found that the fundamental plane becomes a little bit tighter after removing FR I sources, even the slope of $\xi_{\rm X}$ is roughly unchanged.

  We further divide the sample (excluding FR Is) into two subsamples with $L_{\rm X}> L_{\rm X,c}$ (sub-Eddington sources) and $L_{\rm X}< L_{\rm X,c}$ (quiescent sources) respectively, where the fundamental planes are shown in bottom-left and bottom-right panels of Figure 1 respectively. The best fit for sources with $L_{\rm X}> L_{\rm X,c}$ is
  \begin{equation}
  \log L_{\rm R}=0.53^{+0.23}_{-0.19} \log L_{\rm X}+ 0.71^{+0.20}_{-0.16}\log M_{\rm BH}+10.36^{+0.29}_{-0.28},
   \end{equation}
  with  an intrinsic $\sigma_{\rm int}=0.16$ dex. The best fit for sources with $L_{\rm X}\la L_{\rm X,c}$ is
  \begin{equation}
  \log L_{\rm R}=0.55^{+0.17}_{-0.13} \log L_{\rm X}+ 0.77^{+0.14}_{-0.12}\log M_{\rm BH}+9.17^{+0.44}_{-0.44},
   \end{equation}
  with  an intrinsic $\sigma_{\rm int}=0.27$ dex.

  To investigate the radio--X-ray correlation and eliminate the mass effect in AGNs, we further select sources with BH mass in a narrow range but with a broad range of Eddington ratios, which can be used to simulate a single supermassive BH evolving from LH state to quiescent state in a statistical manner. Due to the evident differences in the slope of radio-X-ray correlation for FR Is and other LLAGNs, we explore this issue for these two populations separately (their BH mass distributions are also much different). We select 28 LLAGNs from Table (1), which have BH mass $M_{\rm BH}=10^{8\pm0.4}M_{\sun}$ and Eddington ratios of $L_{2-10\rm keV}/L_{\rm Edd}\sim10^{-8.8}$ to $10^{-3}$. The result is shown in Figure 2, where the faintest sources with $L_{\rm X}\la L_{\rm X,c}$ (solid circles) roughly follow the trend of other sub-Eddington sources. The best-fit linear regression for all sources yields
   \begin{equation}
  \log L_{\rm R}=0.49\pm0.04 \log L_{\rm X} + 17.53\pm1.64,
   \end{equation}
  with a Spearman correlation coefficient of $r=0.75$ and $p=2.41\times10^{-11}$, where the linear regressions were not given for sources  with $L_{\rm X}>L_{\rm X,c}$ and $L_{\rm X}\la L_{\rm X,c}$ respectively due to the narrow range and large scatter of the data in each population. A similar analysis was also done for 14 FR Is with BH mass within $10^{8.8\pm0.4}M_{\sun}$ and $L_{2-10\rm keV}/L_{\rm Edd}$ range from $10^{-7.6}$ to $10^{-3.6}$. The radio--X-ray correlation of these FR Is is shown in Figure 3, and the best fit is
  \begin{equation}
  \log L_{\rm R}=1.27\pm0.10 \log L_{\rm X} - 12.56\pm3.97,
   \end{equation}
  with a Spearman correlation coefficient of $r=0.80$ and $p=3.44\times10^{-13}$.

  \begin{figure}\label{fig2}
   \includegraphics[width=8cm]{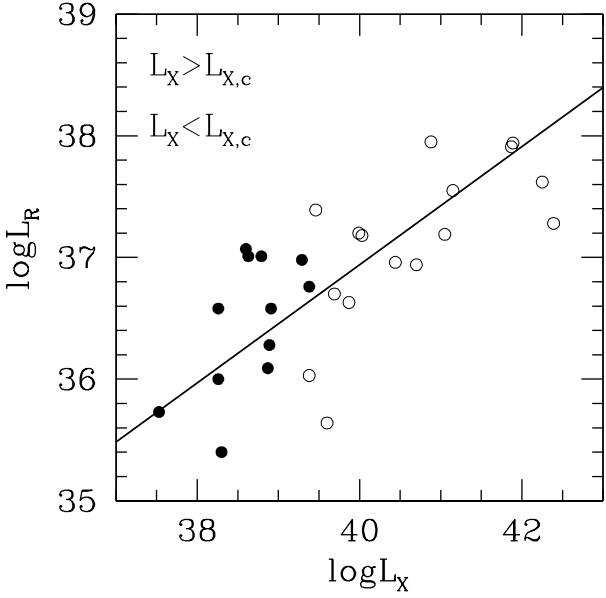}

   \caption{The relation between 5 GHz radio and 2-10 keV X-ray luminosities for LLAGNs with BH mass within $10^{8\pm0.4}M_{\sun}$, where solid and empty circles represent the sources with $L_{\rm X}>L_{\rm X,c}$ and $L_{\rm X}<L_{\rm X,c}$ respectively. The solid line is the best fitting for all sources.   }
  \end{figure}

   \begin{figure}\label{fig3}
   \includegraphics[width=8cm]{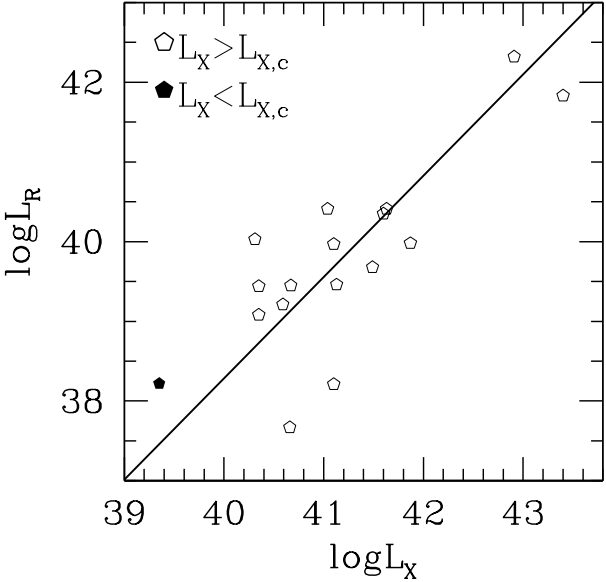}

   \caption{The same as Figure 2, but for FR Is with BH mass within $M_{\rm BH}=10^{8.8\pm0.4}$.   }
  \end{figure}

\section{Conclusion and Discussion}
   In this work, we reinvestigate the radio--X-ray correlation and the fundamental plane of BH activity for a sample of XRBs and LLAGNs from sub-Eddington to quiescent state after including more quiescent BHs. The main results are summarized as follows: 1) the fundamental plane for the quiescent BHs is similar to that of sub-Eddington sources and do not show evident differences; 2) the radio loud AGNs (e.g., FR Is) follow a separate and steeper radio--X-ray correlation comparing with other BH sources; 3) the radio--X-ray correlation can roughly extend to quiescent BHs with $L_{2-10\rm keV}/L_{\rm Edd}\sim10^{-8.8}$ in LLAGNs and $L_{2-10\rm keV}/L_{\rm Edd}\sim10^{-7.6}$ in FR Is, where both subsamples have BH mass in a narrow range that can simulate a single supermassive BH evolving from sub-Eddington to quiescent state. These results are similar to that of XRBs where the quiescent XRBs with $L_{2-10\rm keV}/L_{\rm Edd}<10^{-8.5}$ still follow the radio--X-ray correlation as defined in LH state \citep[][for A0620-00 and XTE J1118+480]{gall06,gall14}.

   The radio--X-ray correlation plays an important role in understanding the physics of BH central engines, which was widely explored in XRBs and AGNs. It is roughly consensus that faint BHs with $L_{\rm bol}/L_{\rm Edd}\la 1\%$ follow a shallower radio--X-ray correlation with $\xi_{\rm X}\sim0.5-0.7$ \citep[e.g.,][]{corb03,gall03,corb13} while brighter BHs with $L_{\rm bol}/L_{\rm Edd}\ga 1\%$ follow a steeper track with $\xi_{\rm X}\sim1.2-1.6$ \citep[e.g.,][]{cori11,dw14,pa15}. However, it is still debatable whether the quiescent BHs still follow the radio--X-ray correlation with $\xi_{\rm X}\sim0.5-0.7$ as defined in LH state of XRBs or not. \cite{yuan09} found the LLAGNs with $L_{\rm X}\la L_{\rm X,c}\sim10^{-6}-10^{-5}L_{\rm Edd}$ follow a steeper radio--X-ray correlation (e.g., $L_{\rm R}\propto L_{\rm X}^{1.22}$), which is roughly consistent with their model prediction in \citet{yc05} where the X-ray emission switches from being ADAF to jet dominated. However, several quiescent XRBs challenge this conclusion where the sources with $L_{\rm X}\la 10^{-8.5}L_{\rm Edd}$ still follow the radio--X-ray correlation as defined by LH state of XRBs very well (e.g., A0620-00, \citealt{gall06} and XTE J1118+480, \citealt{gall14}). We further explore this issue using a large sample of BHs from highly sub-Eddington to sub-Eddington. We don't find that the quiescent BHs with $L_{\rm X}< L_{\rm X,c}$ are different from other sub-Eddington BHs based on the analysis of the radio--X-ray correlation and the fundamental plane. In particular, we also use two subsamples with similar BH mass (FR Is with $M_{\rm BH}=10^{8.8\pm0.4}$ and other LLAGNs with $M_{\rm BH}=10^{8.0\pm0.4}$) to simulate the behavior of a single supermassive BH evolving across from sub-Eddington to quiescent state (e.g., LLAGNs have $L_{\rm X}\sim 10^{-9}-10^{-3}L_{\rm Edd}$ and FR Is have $L_{\rm X}\sim 10^{-7.6}-10^{-3.6}L_{\rm Edd}$). The quiescent BHs still roughly follow that defined by the whole sample. Our results are consistent with that of XRBs \citep[][]{gall06,gall14},but is different from that derived from AGNs \citep{yuan09}. The possible reason is that BH sources observed with large-scale relativistic jets (e.g., FR Is) are included in \citet{yuan09}'s sample, where RL AGNs normally follow a much steeper radio--X-ray correlation regardless of Eddington ratios (see also $L_{\rm R}\propto L_{\rm X}^{1.5}$, \citealt{li08,de11} or FR I sample in this work). The bimodal distribution of radio loudness in AGNs is still an open issue. From the slope of radio--X-ray correlation, we can find that the normal LLAGNs may be similar to LH state of XRBs while RL FR Is with relativistic jets may be intrinsically different. However, it should be noted that the intrinsic correlation may be different comparing with the observed radio--X-ray in RL AGNs, where the Doppler boosting effect is not considered in our work (also \citealt{li08,yuan09,de11}) which may correlate with the luminosities \citep[e.g.,][]{ka14}. Better constraints on the Doppler factors in both XRBs and AGNs are expected to further investigate this issue. It should be also noted that our subsample with similar BH mass is still limited. More sources with similar BH mass and a wide distribution of Eddinton ratio are expected to further test the possible radio--X-ray correlation from quiescent to sub-Eddington BHs.

   The radio emission is normally believed to originate from the jet in both XRBs and AGNs. The radio--X-ray correlation in BH systems suggests the possible coupling of disc and jet, where the X-ray emission dominantly comes from ADAF or corona \citep[but see][for a different opinion]{ma05}. The origin of X-ray emission from sub-Eddington to quiescent BHs may not change (e.g., ADAF or jet) if the radio--X-ray as found in LH state of XRBs can extend to quiescent state. \citet{yc05} predicted that the radio--X-ray correlation will become steeper if $L_{\rm X}\la L_{\rm X,c}$ and the data points will stay below the radio--X-ray correlation as defined in LH state, where the origin of X-ray emission switches from being ADAF to jet dominated. If this is the real case, the observational results suggest that the critical X-ray luminosity may be lower than that reported in \citet{yc05}, where there are still many uncertainties in the ADAF-jet model. In modeling the SED of quiescent BHs with ADAF-jet model, \citet{xie14} suggested that the X-ray emission of V404 Cyg in quiescent state should be dominated by jet based on its X-ray shape \citep[see also][for a couple of quiescent supermassive BHs]{yuan09}, where jet produce power-law X-ray spectrum while the inverse-Compton spectrum from ADAF is normally curved. The soften of X-ray spectrum in quiescent BHs can be explained by both ADAF and jet models separately, where the bremsstrahlung radiation will contribute to X-ray emission \citep[][]{es97} while the synchrotron emission normally produce a power-law X-ray with $\Gamma\sim2$ \citep[e.g.,][]{yuan09,xie14,yang15}. Broadband X-ray observation in quiescent BHs and better dealing with Comptonization in ADAF are expected to further test this issue. 

   The RL AGNs normally follow a steeper radio--X-ray correlation \citep[e.g., FR Is, see also][]{li08,de11}. The physical reason for steeper correlation as found in FR Is is unclear, which may include: 1) the origin of the X-ray emission is different, where the X-ray come from accretion flows in RQ LLAGNs while it is, similar to radio, come from the jet in RL FR Is \citep[e.g.,][]{hard09}; 2) X-ray emission always come from the jet for both RQ AGNs and RL AGNs, but the effect of jet synchrotron cooling is depend on BH mass, where most of the RL FR Is have more massive BHs (e.g., $M_{\rm BH}\ga 10^{8}\sun$) and may follow a steeper radio-X-ray correlation that regulated by synchrotron cooling in the jet \citep[e.g.,][]{plot13}; 3) different accretion-jet properties, where the stronger radio jet in FR Is may be regulated by a rapidly rotating BH while the radio emission of radio weak AGNs is dominated by weak jets or disk winds \citep[][]{wu11}. Better constraints on the origin of X-ray emission and/or the possible correlation between jet and BH spin for more sources may shed light on this issue. Furthermore, a more uniform and larger sample with wider distributions of radio loudness and Eddington ratio is wished to further explore the possible \`fundamental plane'.


 \section*{Acknowledgments}
   We thank HUST astrophysics group for many useful discussions and comments. This work is supported by the NSFC (grants 11103003, 11133005, 11573009 and 11303010) and New Century Excellent Talents in University (NCET-13-0238).

\label{lastpage}

\end{document}